# REVERSALS OF GNEVYSHEV - OHL RULE

Andrey G. Tlatov

*Kislovodsk Mountain Astronomical Station of Pulkovo Observatory, Russia;*

*357700, Gagarina str. 100, Kislovodsk, Russia*

*tlatov@mail.ru*

ABSTRACT

This paper considers the changes of average number of sunspots groups per day, aggregated by activity cycles $G_n$ within the period since 1610 till present time. There is a positive correlation between the parameter $G_n$ in the current cycle and analogous parameter in the following cycle $G_{n+1}$ both for pairs of even and odd cycles and for pairs of odd and even activity cycles. For cycles № 10-21 for a pair of even and odd cycles the relation of parameter $G_n$ corresponds to Gnevyshev - Ohl rule $G_{n+1}^{odd}/G_n^{even}$ >1. But in previous epochs the odd cycles could be less than the preceding even cycles, and it only proves the reversal of Gnevyshev - Ohl rule. In the period after Maunder Minimum for pairs $G_{n+1}^{odd}/G_n^{even}$ there is a long-term variation with the period of about ~21 activity cycles.

## 1. INTRODUCTION

Gnevyshev - Ohl empirical rule (GO) is formulated for a pair of subsequent solar cycles (Gnevyshev & Ohl 1948). There are several concepts of GO rule: a) the amplitude of the even activity cycle is less than the altitude of the following odd cycle; b) the sum of Wolf number in the even activity cycle is less than the sum of the following odd cycle; c) the area under the curve of Wolf numbers correlates to the area under the curve of the odd cycles, and in this case the even activity cycle forms a pair with the following odd cycle (Kopecky 1950;

Hathaway et al. 2002; Nagovitsyn et al. 2009; Ogurtsov & Lindholm 2011). **These** formulation is called the GO1 effect.

If we exclude the anomalous pair of cycles 4-5, then we will see that there is a strong interrelation (R=0.91) between the sums of Wolf numbers for the period of even and odd cycles in the combination "even-odd". For pairs of odd-even cycles this interrelation is weak R=0.50 (Vitinsky et al. 1986; Wilson 1988; Ogurtsov & Lindholm 2011). Here, this formulation is called the GO2 effect.

On the basis of Wolf numbers this is right for the cycles beginning with the $10^{th}$ activity cycle, but it's not satisfactory for the pairs of 4/5 and 8/9 (Gnevyshev & Ohl 1948, Wilson 1988; Hathaway 2010).

Usually, for checking GO rule one uses a sequence of Wolf numbers reconstructed by R. Wolf since 1748 (Wolf 1861). However, this sequence has a sufficient noise due to difficult taking account of small sunspots without taking into consideration other kinds of observations (Hoyt & Schatten 1998; Svalgaard 2012). Having based oneself on additional data, Hoyt & Schatten (1994) offered index of group sunspot numbers, reconstructed by them for the period from 1610 till 1995. The index of group sunspot numbers gives the best coefficient of correlation between even and following odd activity cycle, in comparison with Wolf numbers (Hathaway et al. 2002). In the paper of Mursula et al. (2001) it is offered to use index of sunspots groups for checking GO1 rule in the following way $I_{GO}(k) = 1/132 \sum_{j=J(k)}^{J(k+1)-1} Gm$, where Gm- average monthly index of group sunspot numbers, J(k)- the month of the beginning of the cycle *k*, and the invariable 1/132 is introduced for scaling obtained index towards standard sunspots. The authors showed that during the period since 1725 till 1782 the correlation of even activity cycles turn out to be higher than that of the odd cycles. To eliminate this discrepancy, they offered a hypothesis, according to which one solar cycle was lost in the beginning of Dalton minimum during the 1790s (Usoski et al. 2001;

Usoskin et al. 2009). The index $I_{GO}(k)$ offered by the authors (Mursula et al. 2001) is analogous to the sum of sunspots per cycle. But if there is a considerable gap in observation days, it's difficult to calculate the sum of sunspots per cycle, as well as to define the amplitude of a cycle.

This paper offers to use the index of sunspots groups based not on the summing of number of sunspots groups per cycle, but on defining an average number of sunspots groups per day during the activity cycle.

## 2. DATA AND ANALYSIS

For characterizing activity cycles we can use an average number of sunspots groups per day within the cycle: $G_n = \sum_{T_k}^{T_{k+1}} Rg / Nd$, where Rg- an amount of groups per current day, *Nd*- a quantity of days with observations within the cycle n, $T_n$- the moment of beginning of the cycle n. The basic data (Hoyt et al. 1994) contain the information of different observatories and researchers concerning the number of sunspots groups per day (ftp://ftp.ngdc.noaa.gov/STP/SOLAR_DATA/SUNSPOT_NUMBERS/GROUP_SUNSPOT_NUMBERS/alldata.txt). Within this analysis, when there were several observations per day, the one that had the largest amount of sunspots groups was chosen. The results were checked by means of data, containing interpolated daily values. The moments of the beginning of cycles $T_n$ also were taken from the web-site NGDC.

Table 1 reflects the magnitudes of the parameter $G_n$ from the period since 1610 till present. The values for cycles 23, 24 were received due to calculating the amount of groups according to the sunspots data base USAF/NOAA (http://solarscience.msfc.nasa.gov/greenwch.shtml). The data concerning Wolf numbers exist since 1749 (cycle № 0). There is a positive correlation between the index $G_n$ and the amplitude of the cycle in Wolf numbers: $W_n^{max} = 45.0(12)+19.0(3)G_n$ ; R=0.8.

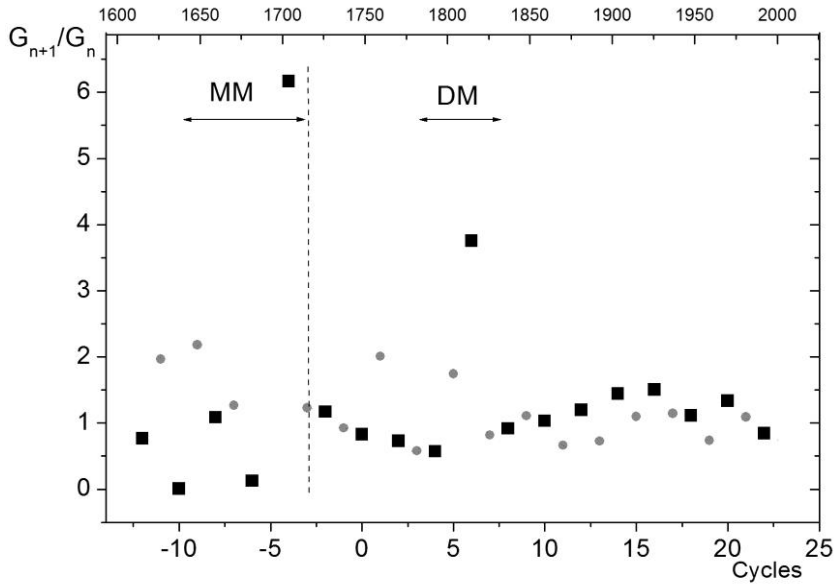

**Figure 1**. The ratio of average daily number of sunspots groups in neighboring cycles $G_{n+1}/G_n$. Squares indicate at pairs of even/ odd cycles, circles show odd/ even. Positions of Maunder Minimum (MM) Dalton Minimum (DM) are presented.

Fig. 1 performs time change ratio of sunspots groups index in the following cycle $G_{n+1}$, towards the previous $G_n$, for cycles N -12 : 23. Pairs $G_{n+1}^{odd}/G_n^{even}$ and $G_{n+1}^{even}/G_n^{odd}$ are presented by different symbols. After the ending of Maunder Minimum (1645-1715), beginning from cycle № -2, for the pairs of even and the following odd cycle one can point out regularity of values $G_{n+1}^{odd}/G_n^{even}$ by way of long-term modulation. The pair of 6-7 cycles, falling within Dalton Minimum is an exception.

The average quantity of sunspots groups in the following cycle $G_{n+1}$ is linked with the quantity of sunspots groups in previous cycle $G_n$. Figure 2 shows the relation of $G_{n+1}^{odd}$ in function from the average number of sunspots groups in the preceding cycle $G_n^{even}$ for pairs of even –odd cycles. The proportion between indexes $G_n$ in such pairs has a positive correlation $G_{n+1}^{odd} = 0.37(0.47) + 0.93(0.16) G_n^{even}$   R=0.82, that corresponds to GO2 rule. At the same time such a correlation exists for pairs of odd-even cycles (Fig. 3). The correlation between indexes for these pairs is  $G_{n+1}^{even} = 0.39(0.4) + 0.82(0.08) G_n^{odd}$,   R=0.91.

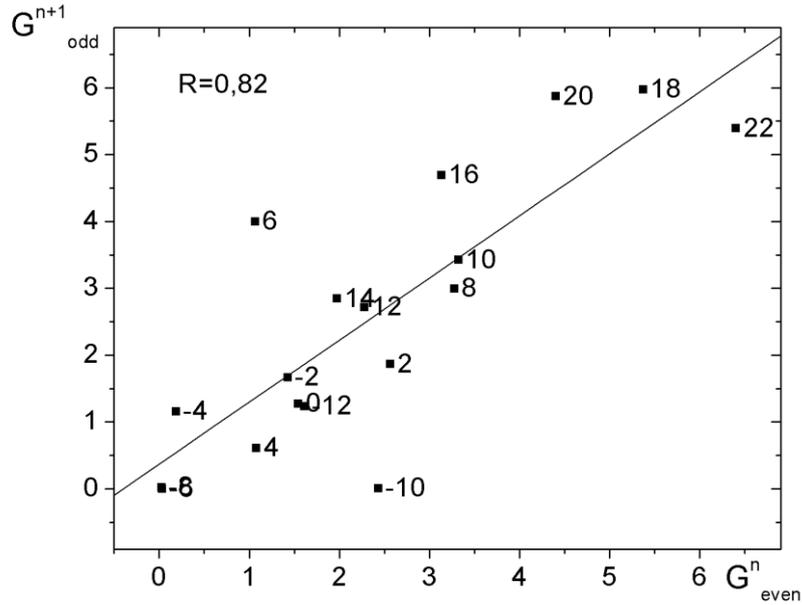

**Figure 2**. The average number of sunspot groups in the current cycle $G_n^{even}$ compared with the number of sunspots groups in the following cycle $G_{n+1}^{odd}$ for pairs of even-odd cycles.

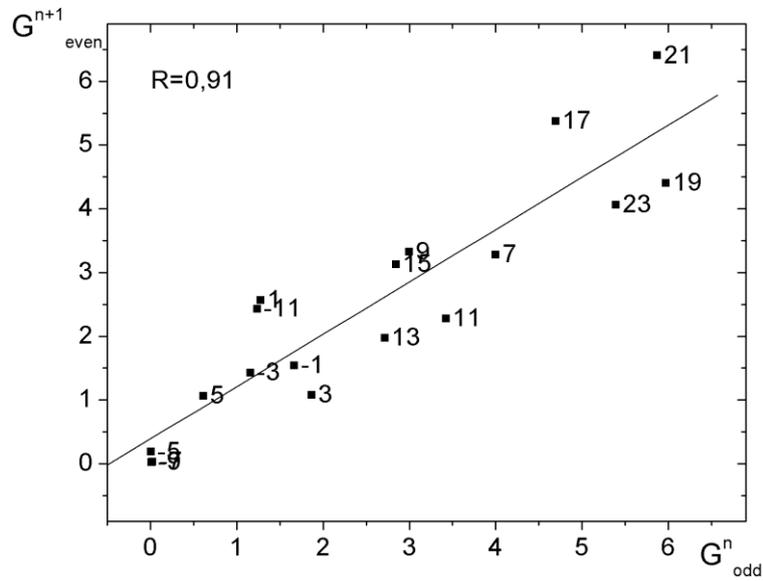

Figure 3. Just like in Fig. 2, but for pairs of odd $G_n^{odd}$ and the following even activity cycles $G_{n+1}^{even}$.

Table 1 shows that during the period № 10-21 the average number of groups in odd cycles were higher than in preceding even cycles and the relation $G_{n+1}^{odd}/G_n^{even}$ corresponds to the standard formulations of GO1 rule, but is violated in pair of 22-23 cycles. Figure 4 represents the relation $G_{n+1}^{odd}/G_n^{even}$ in period after Maunder Minimum. All the pairs of cycles except from cycles 6-7 are within the range of values $0.5 \div 1.5$. Starting from the cycle -2 the relation $G_{n+1}^{odd}/G_n^{even}$ has a smooth envelope. As a comparison, the diagram shows the sinusoid with the period t=21 cycles and the amplitude a=0.45 : $f(t) = 0.5 + 0.45 \cdot \sin(2\pi n/21)$. Standard deviation error in ratio $G_{n+1}^{odd}/G_n^{even}$ from envelope curve amounted to $\sigma = 0.12$.

## 3. DISCUSSION AND CONCLUSIONS

Research of the GO rule can give important information about the nature of the solar periodicity, un particular, concerning the possible fossil solar magnetic field, with which one usually connects this effect (Bravo & Stewart 1995; Charbonneau 2009). Some authors think that the regularity when even cycles are less intensive, than the following odd ones, has a constant character (Usoskin et al. 2001; Nagovitsyn et al. 2009). However, the pair of 22 -23 activity cycles obviously shows violation of this rule (Hathaway 2010). Therefore, possibly, this rule had its reversals in previous centuries.

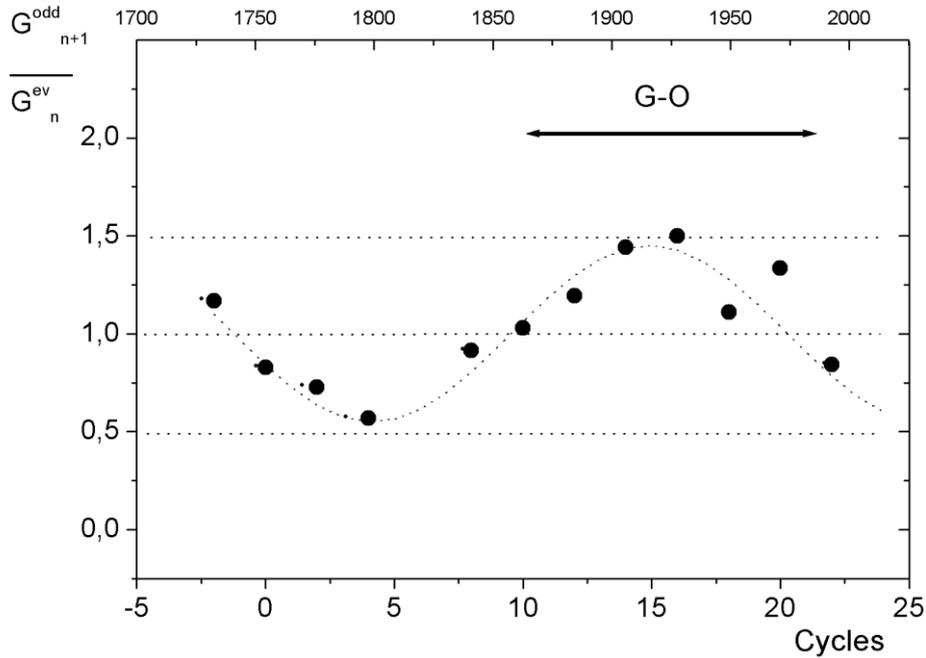

**Fig. 4.** Relation of average daily amount of sunspots groups in odd cycle to analogous value in the preceding in even cycle $G_{n+1}^{odd}/G_n^{even}$. An envelope and a line are drawn, and a line where this relation is equal to 1.

To test GO1 rule, index of average number of sunspots groups in cycle $G_n$ was offered in this paper. We discovered that the correlation $G_{n+1}^{odd}/G_n^{even}$ in cycles № 10-21 in pairs of even-odd cycles was more than 1, while the pair 22-23 showed less than 1, and it corresponds to the results, obtained by means of Wolf numbers (Hathaway 2010). A strong interrelation of $G_n$ parameters was found in pairs of even-odd cycles for all the period under consideration, cycles № -12 ÷ 23 (R=0.82, Fig.2 ). These results correspond to the standard definitions of GO1 and GO2 rule and its exception for pair 22-23, and it indicates at validity of using parameter $G_n$ in testing GO rule.

At the same time, a high positive correlation (R=0.91) was established in pairs of even-odd activity cycles (Fig. 3). Usage of index $G_n$ allowed to single out a long-term envelop curve (dependence) for values $G_{n+1}^{odd}/G_n^{even}$ after Maunder Minimum (Fig. 1). The changes are close to long-term variations with the period of order ~21 cycle (Fig. 4) or about 230 years. Only 6-7 cycles are an exception, because it takes place in Dalton minimum. Some authors (Vitinsky et al. 1986; Mursula et al. 2001) came to the conclusion that apparition of the 22-year periodicity period disappeared in the period when the level of solar activity changed quickly, for instance, during restoring of activity after Maunder Minimum, or closer to the Dalton minimum. Long-term cyclicity with the period about ~200-220 years had been found by means of reconstructing solar activity according to radioisotope data before (Mordvinov & Kramynin, 2010; Abreu et al. 2012). For period of 1750-1840 the value of correlation in pars even-odd cycles $G_{n+1}^{odd}/G_n^{even}$ was less than 1. It gives evidence that GO1 rule can reverse within long intervals, that is – even cycles can be weaker than the following odd cycles. Duration of epochs, when $G_{n+1}^{odd}/G_n^{even}>1$ and $G_{n+1}^{odd}/G_n^{even}<1$ are approximately equal, and reversing takes place during secular activity minimum (Fig. 4). It's possible to expect that the following activity cycles will develop within the reversed GO1 rule.

Presumably, the violation in 22 year cycles has a periodic character, during which the Sun changes its cycle mode. As a rule, one can observe minimums of century variations of solar activity in the process. To explain this, we can assume that in the long-term periods there is a permanent solar magnetic field which can also reverse, and it reverses the sequency of 22 year cycles. The reason of appearing of such a permanent field can be the so-called "magnetic memory" under the bottom generation zone (Tlatov 1996). This field appears during averaging out magnetic field of several subsequent cycles, having different direction of poloidal field, thus ensuring the relation $G_{n+1}/G_n$ higher (lower) than 1 during long-term periods (Figure 1). The positive correlation between the preceding and the following cycle $G_{n+1}^{odd}/G_n^{even}$ and $G_{n+1}^{even}/G_n^{odd}$ (Figure 2, 3), as well as changes with long-term period, which depicts Figure 1, count in favor of this hypothesis.

Therefore, the violation of GO rule in 22-23 activity cycles can be a sign of changing in the character of periodicity period and long-term reversal of GO rule in the following activity cycles. GO rule, established and right for cycles 10-21, is a part of long-period inequality of solar activity.

The work was work was supported by the Russian Foundation for Basic Research (RFBR), the Russian Academy of Sciences, and the Program to Support Leading Scienti_c Schools by the Russian Federal Agency for Science and Innovations.

# REFERENCES

Abreu, J. A., Beer, J., Ferriz-Mas, A., McCracken, K. G. & Steinhilber, F., 2012, A&A, 548, A88.

Bravo, S., & Stewart, G. 1995, ApJ, 446, 431.

Charbonneau, P. 2005, Living Rev. Sol. Phys., 2, 2 (URL cited on 2009.20.01: http://www.livingreviews.org/lrsp-2005-2).

Gnevyshev, M. N., & Ohl, A. I. 1948, Astron. Zh., 25, 18.

Hathaway, D. 2010, Living Rev. Sol. Phys., 7, 1 (URL cited on: http://www.livingreviews.org/lrsp-2010-7).

Hathaway, R. M., Wilson, R. M., & Reichmann, E. J. 2002, Solar Phys., 211, 357.

Hoyt, D. V., Schatten, K. H., & Nesmes-Ribes, E. 1994, Geophys. Res. Lett., 21, 2067

Hoyt, D., & Schatten, K. 1998, Sol. Phys., 181, 491.

Kopecky, M. 1950, Bull. Astron. Instit. Czech., 2, 14.

Mordvinov, A.V. & Kramynin, A.P., 2010, Sol. Phys., 264, 269.

Mursula, K., Usoskin, I. G., & Kovaltsov, G. A. 2001, Sol. Phys., 198, 51.

Ogurtsov, M. & Lindholm, M. 2011, ISRN Astronomy and Astrophysics, 2011, ID 640817, doi:10.5402/2011/640817.

Nagovitsyn, Yu. A.; Nagovitsyna, E. Yu., & Makarova, V. V. 2009, Astr. Lett., 35, 564.

Svalgaard, L. 2012, IAU Symp., 286, 27.

Tlatov, A. G. 1996, Radioph. and Quant. Elect., 39, 794.

Usoskin, I. G.. Mursula, K.; Arlt, R. & Kovaltsov, G. A. 2009, ApJL, 700, L154.

Usoskin, I. G.; Mursula, K. & Kovaltsov, G. A. 2001, A& A, 370, L31.

Vitinsky, Yu. I., Kopecky, M. & Kuklin, G. V. 1986, The statistics of sunspots (Statistika pjatnoobrazovatelnoj dejatelnosti solntsa), Moscow: Nauka, 397 pp. (In Russian).

Wilson, R. M. 1988, Sol. Phys., 117, 269.

Wolf, R. 1861, MNRAS, 21, 77.

**Table 1.** Index of average daily number of sunspots groups Gn in activity cycles, Confidence-interval $\mu=\sigma N^{1/2}$, The number of observation days in the activity cycle N, The relation of average number of groups in the following cycle to the previous $G_{n+1}/G_n$, The moment of activity cycles' minimum Tmin.

| № cycle | $G_n$ | μ | N | $G_{n+1}/G_n$ | Tmin |
|---|---|---|---|---|---|
| -12 | 1.614 | 0.071 | 1022 | 0.766 | 1610.8 |
| -11 | 1.236 | 0.039 | 1469 | 1.966 | 1619 |
| -10 | 2.43 | 0.069 | 1634 | 0.005 | 1634 |
| -9 | 0.011 | 0.003 | 3209 | 2.182 | 1645 |
| -8 | 0.024 | 0.002 | 3926 | 1.083 | 1655 |
| -7 | 0.026 | 0.002 | 4270 | 1.269 | 1666 |
| -6 | 0.033 | 0.003 | 3629 | 0.121 | 1679.5 |
| -5 | 0.004 | 0.002 | 3378 | -- | 1689.5 |
| -4 | 0.188 | 0.008 | 3800 | 6.165 | 1698 |
| -3 | 1.159 | 0.05 | 2772 | 1.23 | 1712 |
| -2 | 1.426 | 0.072 | 819 | 1.169 | 1723.5 |
| -1 | 1.667 | 0.137 | 150 | 0.926 | 1734 |
| 0 | 1.544 | 0.072 | 612 | 0.827 | 1745 |
| 1 | 1.277 | 0.041 | 1317 | 2.009 | 1755.2 |
| 2 | 2.565 | 0.054 | 1388 | 0.727 | 1766.5 |
| 3 | 1.866 | 0.076 | 614 | 0.578 | 1775.5 |
| 4 | 1.079 | 0.054 | 693 | 0.567 | 1784.7 |
| 5 | 0.612 | 0.025 | 2695 | 1.742 | 1798.3 |
| 6 | 1.066 | 0.024 | 3007 | 3.753 | 1810.6 |
| 7 | 4.001 | 0.086 | 3522 | 0.82 | 1823.3 |
| 8 | 3.279 | 0.058 | 2539 | 0.914 | 1833.9 |
| 9 | 2.997 | 0.041 | 4337 | 1.109 | 1843.5 |
| 10 | 3.323 | 0.043 | 4065 | 1.03 | 1856 |
| 11 | 3.424 | 0.048 | 4234 | 0.665 | 1867.2 |
| 12 | 2.277 | 0.032 | 3908 | 1.193 | 1878.9 |
| 13 | 2.717 | 0.037 | 4419 | 0.727 | 1889.6 |
| 14 | 1.975 | 0.029 | 4346 | 1.441 | 1901.7 |
| 15 | 2.846 | 0.044 | 3652 | 1.1 | 1913.6 |
| 16 | 3.131 | 0.042 | 3726 | 1.5 | 1923.6 |
| 17 | 4.696 | 0.06 | 3798 | 1.144 | 1933.8 |
| 18 | 5.374 | 0.069 | 3689 | 1.112 | 1944.2 |
| 19 | 5.974 | 0.08 | 3872 | 0.737 | 1954.3 |
| 20 | 4.402 | 0.048 | 4236 | 1.334 | 1964.9 |
| 21 | 5.871 | 0.076 | 3762 | 1.091 | 1976.5 |
| 22 | 6.405 | 0.09 | 3360 | 0.719 | 1986.8 |
| 23 | 5.39 | 0.057 | 4419 | 0.79 | 1996.9 |
| 24 | 4.28 | 0.076 | 1016 | | 2009 |